# Enhanced colloidal transport in twisted magnetic patterns

Nico C. X. Stuhlmüller[1], Thomas M. Fischer[2] & Daniel de las Heras[1 ✉]

Bilayers of two-dimensional materials twisted at specific angles can exhibit exceptional properties such as the occurrence of unconventional superconductivity in twisted graphene. We demonstrate here that novel phenomena in twisted materials emerges also in particle-based classical systems. We study the transport of magnetic colloidal particles driven by a drift force and located between two twisted periodic magnetic patterns with either hexagonal or square symmetry. The magnetic potential generated by patterns twisted at specific magic angles develops flat channels, which increase the mobility of the colloidal particles compared to that in single patterns. We characterize the effect of the temperature and that of the magnitude of the drift force on the colloidal mobility. The transport is more enhanced in square than in hexagonal twisted patterns. Our work extends twistronics to classical soft matter systems with potential applications to lab-on-a-chip devices.

[1] Theoretische Physik II, Physikalisches Institut, Universität Bayreuth, D-95440 Bayreuth, Germany. [2] Experimatalphysik X, Physikalisches Institut, Universität Bayreuth, D-95440 Bayreuth, Germany. ✉email: delasheras.daniel@gmail.com

The emerging field of twistronics[1] studies the properties of bilayers of two-dimensional materials that are rotated relative to each other by a twist angle. Twisted bilayers generically create a quasiperiodic moiré pattern. However, for specific twist angles, the pattern is periodic with a super unit cell that is a multiple of the primitive unit cell of a monolayer. New properties, not present in the individual monolayers, can emerge in the resulting moiré superlattices[2–4]. These include superconductivity[5,6], ferromagnetism[7], and correlated insulating states[8–11] in twisted bilayer graphene. The formation of moiré patterns in twisted materials also affects the properties of both light[12–16] and acoustic waves[17–20].

We extend here twistronics to a classical, particle-based, system made of magnetic colloidal particles that are located between two periodic magnetic patterns and are driven by a weak drift force. The patterns are twisted at a small angle. For specific magic angles, a partial destructive interference between the magnetic fields of the patterns generates flat channels in the total magnetic potential which results in enhanced long-range anisotropic colloidal transport. We study with Brownian dynamics simulations the effect of the twist angle, the temperature, and the magnitude of the drift force on the mobility of the particles for both square and hexagonal twisted patterns. Our results and conclusions may apply to other twisted materials and constitute the basis for novel lab-on-a-chip applications.

## Results

**Setup**. We study the motion of paramagnetic colloidal particles confined to the middle plane between two parallel periodic magnetic patterns that are separated by a distance $\Delta$, see Fig. 1a. We consider both square and hexagonal periodic lattices with regions of positive and negative magnetization normal to the pattern. The patterns are twisted by an angle $\alpha$ and shifted by half a unit lattice vector.

A uniform external magnetic field $\mathbf{H}_{ext}$, much stronger than the pattern fields ($\mathbf{H}_{p,i}$ with $i = 1, 2$) points normal to the patterns. At vertical distances comparable or larger than the size of the unit cell, i.e., $\Delta \gtrsim a$ with $a$ the magnitude of the lattice vectors, the total magnetic potential is dominated by the coupling between external and pattern fields: $V_{mag} \propto -\sum_i \mathbf{H}_{ext} \cdot \mathbf{H}_{p,i}$. Hence, only the components of the pattern fields normal to the patterns contribute to the potential. The magnetic potential of single square and hexagonal patterns in presence of $\mathbf{H}_{ext}$ is shown in Fig. 1b. Details about the calculation of $V_{mag}$ are provided in Methods.

Using single patterns it is possible to transport the particles via modulation loops of the orientation of the external field[21,22]. The loops are closed such that the orientation returns to its initial value after one loop. Loops that wind around special directions of the external field transport the particles by one unit cell after completion of the loop. The transport is topologically protected since the precise shape of the loop is irrelevant, only the winding numbers around the special directions determine the transport. Here, we explore a different type of transport. We keep the external field constant in time and apply a uniform static external drift force, $\mathbf{f}_d$, in the plane parallel to the patterns, see Fig. 1a. We calculate the particle trajectories using overdamped Brownian dynamics simulations. Hence, the equation of motion for a single particle reads:

$$\xi \dot{\mathbf{r}} = -\nabla V_{mag}(\mathbf{r}) + \mathbf{f}_d + \boldsymbol{\eta}, \quad (1)$$

where $\xi$ is the friction coefficient against the (implicit) solvent, $\dot{\mathbf{r}}$ is the time derivative of the position vector, and $\boldsymbol{\eta}$ is the delta-correlated random thermal force with standard deviation and amplitude given by the fluctuation-dissipation theorem. We work in units of the magnitude of a lattice vector $a$, the energy parameter of the magnetic potential $\varepsilon$ (see Methods), and the friction coefficient $\xi$. The intrinsic time-scale is $\tau = \xi a^2/\varepsilon$ and absolute temperature $T$ is measured in reduced units $k_B T/\varepsilon$ with $k_B$ the Boltzmann constant.

The amplitude of the drift force is small compared to the magnetic forces. Hence, colloidal transport is not possible above a single pattern lacking flat channels. If the temperature is not high enough to overcome the potential barriers, the particles simply diffuse near the minima of the magnetic potential, see characteristic trajectories in Fig. 1b. The situation is different for the case of twisted patterns.

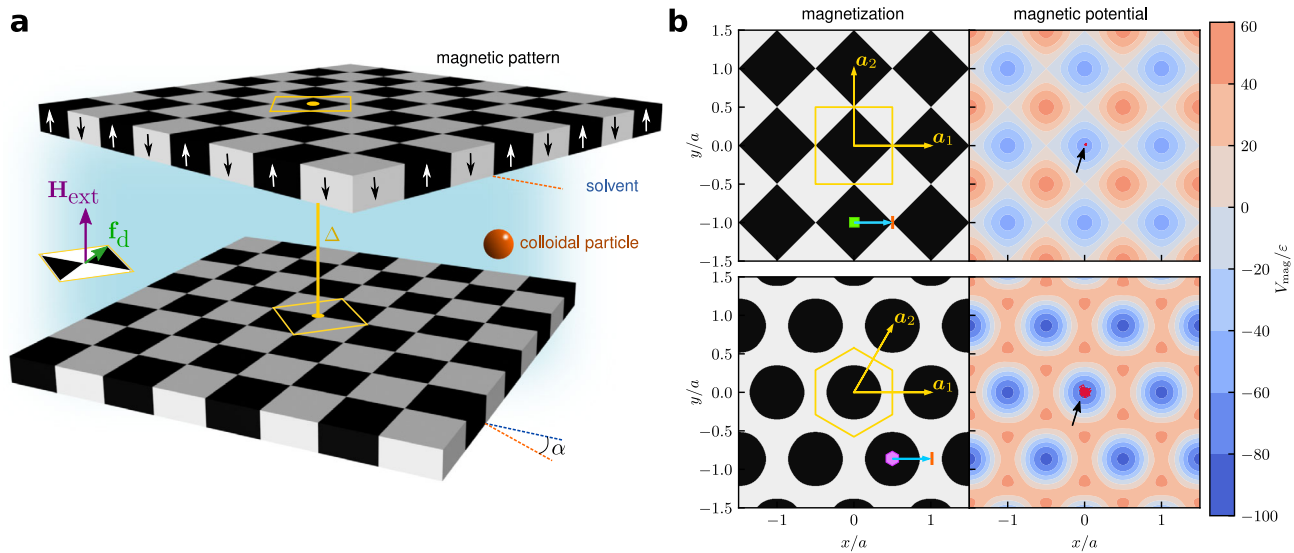

**Fig. 1 Setup. a** Paramagnetic colloidal particles immersed in a solvent are placed between two periodic and parallel magnetic patterns twisted by an angle $\alpha$ and shifted by half a lattice vector. The magnetization of the patterns is indicated by the white and black arrows. A uniform external magnetic field $\mathbf{H}_{ext}$ points normal to the patterns. A drift force $\mathbf{f}_d$ points in the plane of the patterns. **b** Magnetization and magnetic potential $V_{mag}$ of single square and hexagonal patterns. A unit cell together with the lattice vectors $\mathbf{a}_i$, $i = 1, 2$ are indicated in yellow. The shift vectors applied to the twisted system, $\mathbf{a}_1/2$, are represented in blue. The shift vectors connect points with fourfold (green square) or sixfold (violet hexagon) symmetry to points with twofold symmetry (orange rectangles). The trajectories of a particle above only one of the patterns and located in the central unit cell are plotted in red and indicated by black arrows (drift force $f_d a/\varepsilon = 10$). The temperature is set to $k_B T/\varepsilon = 0.01$ (square pattern) and $k_B T/\varepsilon = 1$ (hexagonal pattern).

**Enhanced transport in twisted patterns.** At specific twist angles, the so-called magic angles, a periodic moiré pattern with supercells of size roughly given by $a/(2\sin(\alpha/2))$ develops in the magnetic potential, see Fig. 2 and Supplementary Fig. 1. At a magic angle $\alpha_m^p$ with $p = \{sq, hex\}$ for square and hexagonal patterns, respectively, a lattice site of the twisted pattern coincides with another lattice site of the other pattern (see Methods). At the center of the supercells $V_{mag}$ resembles that of the underlying square or hexagonal patterns. However, near the edges of a supercell, a change towards a twofold symmetry (stripes) occurs. The stripes in $V_{mag}$ are almost flat compared to the inner regions due to a partial destructive interference of the magnetic field of both patterns. A small drift force is therefore able to push the particles along the edges of the supercells, while its effect is negligible for particles inside the supercells, see particle trajectories in Fig. 2 and Supplementary Movies 1, 2.

The patterns are shifted by half the first lattice vector such that points with different rotational symmetries (see Fig. 1b) coincide in the combined system. The shift creates a twofold symmetric point of the magnetic potential at the origin (axis of rotation). This is only possible by shifting the patterns by half a lattice vector and it maximizes the destructive interference between the fields of both patterns. Any lattice vector can be used since the resulting magnetic potentials are the same up to a global rotation. The shift generates a combined pattern which is anisotropic. The curvature of the magnetic potential at the edges of the supercell is either negative ($V_{mag}$ is minimum) or positive ($V_{mag}$ is maximum), see Fig. 2 and Supplementary Fig. 1. For weak drift forces, the particles are transported only along the edges for which the potential is a minimum. In analogy to the flat bands that occur in reciprocal space in twisted graphene[3], we call these edges in real space flat channels. It is worth mentioning that there exists a correspondence between real space and reciprocal space, and that flat channels in real space, similar to those in Fig. 2, also occur in twisted bilayers of 2D materials[23]. At magic twist angles, the flat channels of neighboring cells are connected and the particles can be transported over macroscopic distances.

The roughness of the potential increases progressively from the edges towards the center of the supercell. At multiple distances of $a$ from an edge with a flat channel and parallel to it there exist secondary channels. Secondary channels also occur parallel to the edges for which $V_{mag}$ is maximum. There, the first two secondary channels are located at a distance of $a/2$ from the edge. Along the secondary channels, the magnetic potential is still flat enough to support transport for either strong driving or high temperature as demonstrated below. The smaller the magic angle, the larger the supercell is and more secondary channels are sufficiently flat to transport particles. In addition, the potential along both the secondary and the flat channels gets flattened by decreasing the twist angle. For all magic angles, the flat channel (located at the edges of the supercell) is always the flattest one and requires therefore the minimal drift force to transport particles.

**Critical drift force.** To investigate the minimal (critical) drift force $f_c$ required to achieve macroscopic transport, we point the drift force along the average direction between two consecutive flat channels, see Fig. 2 and Methods, and set the temperature to zero such that the diffusive motion due to thermal fluctuations does not hinder the phenomenology. Then we measure the mobility of the colloidal particles, $\mu$, defined as the average velocity divided by the amplitude of the drift force, see details in Methods. The mobility vanishes for small drift forces, increases abruptly at a given critical value $f_c$, and it saturates for strong drift forces, see illustrative examples in Fig. 3a. The value and the behavior of the critical drift force depend on both the type of pattern and the magic angle $\alpha_m^p$, see Fig. 3b.

At the corners of a supercell (intersection between two edges), the magnetic potential has more structure than in the center of the edges, see Fig. 2. Hence, for both hexagonal and square twisted patterns, the transport along the edges of the supercells requires weaker drift forces than the transport over the corners. The particles spend a significant amount of time crossing the corners. This effect can be observed in Supplementary Movies 1 and 2 which show the particle dynamics in square and hexagonal twisted patterns, respectively. Plots of the crossing time at the corners for different temperatures and magnitudes of the drift force are shown in Supplementary Fig. 2. In square twisted patterns the height of the magnetic potential near the corners increases with the twist angle, and also the potential at the edges becomes rougher. These two effects cause the critical force to increase monotonically with the twist angle, Fig. 3b. In the limit $\alpha_m^{sq} \to 0$ the critical force vanishes and macroscopic transport occurs for an infinitesimally small drift force. Note that in the limit of vanishing twist it is also necessary to have both patterns in order to create a destructive interference along the direction of the shift.

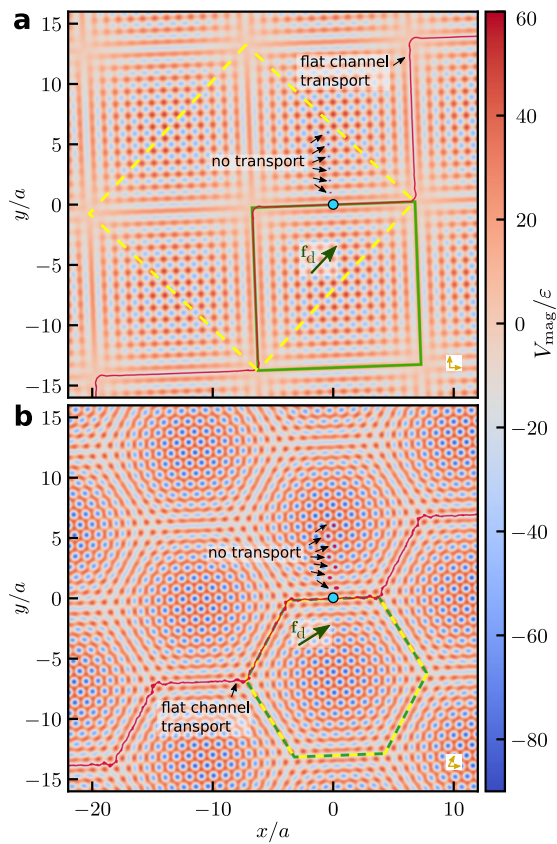

**Fig. 2 Twisted patterns.** Magnetic potential, $V_{mag}$, in square (**a**) and hexagonal (**b**) twisted patterns. The patterns are twisted at a magic angle of $\alpha_m^{sq} \approx 4.24°$ in (**a**) and $\alpha_m^{hex} \approx 4.41°$ in (**b**) around the axis normal to the patterns that passes through the origin (blue circle). The lattice vectors, $\mathbf{a}_1$ and $\mathbf{a}_2$ with $\mathbf{a}_1$ parallel to the x-axis, of the untwisted pattern are represented with yellow arrows. The untwisted pattern is shifted by $\mathbf{a}_1/2$. A supercell (green solid line) and a unit cell (dashed yellow line) of the twisted patterns is highlighted in each pattern together with trajectories followed by both, particles transported via flat channels and particles stuck inside the supercells, as indicated. The magnitude of the drift force is $f_d a/\varepsilon = 10$ and its direction is indicated by a green arrow. The temperature is $k_B T/\varepsilon = 0.01$ in (**a**) and $k_B T/\varepsilon = 0.8$ in (**b**).

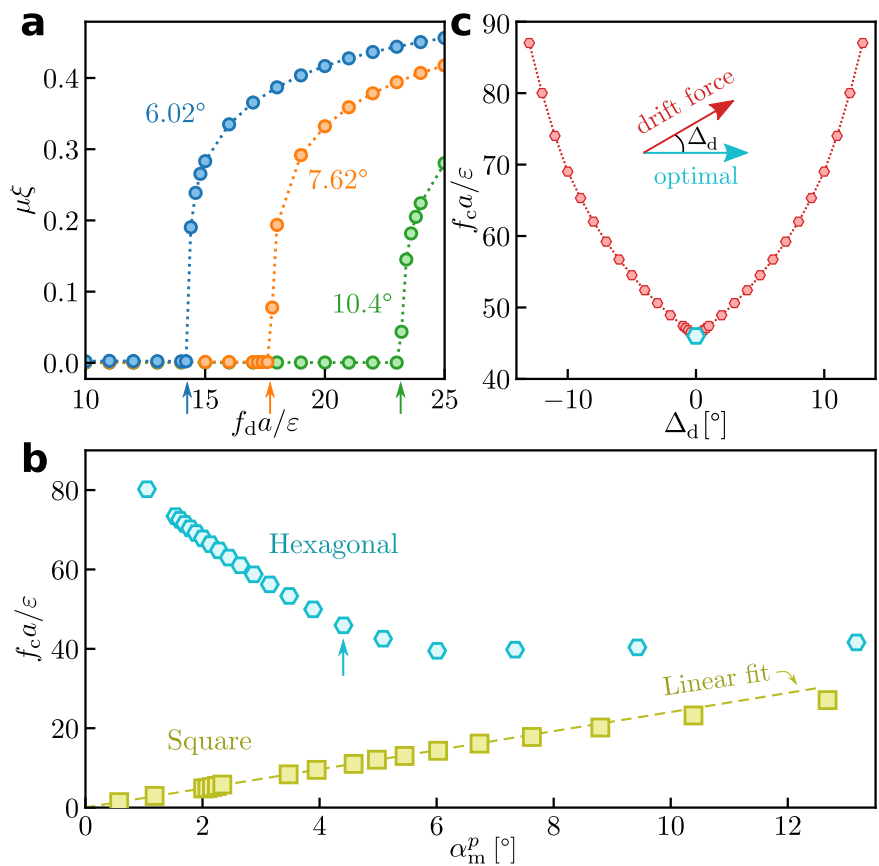

**Fig. 3 Critical force. a** Mobility $\mu$ in twisted square patterns as a function of the magnitude of the drift force $f_d$ for three magic angles. The vertical arrows indicate the value of the critical forces for each magic angle. **b** Magnitude of the critical force $f_c$ required to transport particles at $T = 0$ as a function of the magic angle $\alpha_m^p$ in twisted square (green) and hexagonal (blue) patterns. The dashed green line is a linear fit for magic angles $\alpha_m^{sq} < 4°$ in twisted square patterns. The critical force extrapolates to zero in the limit $\alpha_m^{sq} \to 0$. **c** Magnitude of the critical force $f_c$ as a function of the direction of the drift force $\Delta_d$ (measured as the angle formed by the drift force and the optimal drift force). Data at $T = 0$ for hexagonal patterns twisted at a magic angle $\alpha_m^{hex} \approx 4.41$, indicated with a vertical arrow in (**b**). Dotted lines in panels (**a**) and (**c**) are guides for the eye.

In twisted hexagonal patterns, the potential at the edges of the supercells also gets rougher by increasing the twist angle. However, in contrast to twisted square patterns, the potential at the corners of a supercell becomes flattened by increasing the magic angle. The balance between these opposing effects generates a non-monotonic critical force in hexagonal twisted patterns, Fig. 3b. By increasing the magic angle, the critical force first decreases, then it reaches a minimum at $\alpha_m^{hex} \approx 6.01°$, and finally it increases again.

**Direction of the drift force**. To investigate the effect of the direction of the drift force, we fix the twist angle to a magic angle in hexagonal patterns and calculate the critical force as a function of $\Delta_d$, which is the angle between the drift force and the average direction between two consecutive flat channels. Any deviation of the drift force from the average direction between two consecutive flat channels increases the critical force, see Fig. 3c. For deviations larger than those shown in Fig. 3c transport along the flat channels is no longer possible. Instead, a drift force ($\sim 200\, \varepsilon/a$) stronger than the magnetic forces of the single patterns is then required to transport the particles that no longer follow the flat channels. We also show in Supplementary Fig. 3 the critical force as a function of the magic angle for different directions of the drift force.

**Finite temperature**. We next characterize the effect of Brownian motion on colloidal transport. We show in Fig. 4 the dynamical phase diagram of the colloidal mobility in the plane of temperature and magnitude of the drift force for patterns twisted at a magic angle.

The highest mobility occurs for strong forces and low temperatures. By increasing the temperature at a constant magnitude of the drift force, the mobility first decreases and then increases again. This effect is more prominent in the square case, Fig. 4a, although it also occurs in hexagonal twisted patterns, Fig. 4b. The first minimum in the mobility is caused by particles getting scattered off the central flat channel due to Brownian motion. (A conceptually related Pomeranchuk effect in twisted graphene[24,25] in which increasing temperature induces a metal-insulator transition has been recently observed.) The secondary channel does not allow for macroscopic transport in a given range of temperatures and force amplitudes. However, a further increase in the temperature allows transport along the secondary channel (larger thermal fluctuations permit the particles to cross the potential barriers) and the mobility increases again. At even higher temperatures the mobility decreases again since particles get scattered into the next secondary channel. This oscillatory behavior in the mobility continues until the thermal energy is large enough compared to the magnetic potential energy and the transport becomes diffusive.

A similar argument explains also the negative differential mobility observed by increasing the magnitude of the drift force at finite temperature ($T > 0$). First $\mu$ increases, as expected, and then it decreases. If a particle leaves the flat channel at the corners of a supercell (due to Brownian motion), it is driven away from the corner faster for stronger $f_d$. This decreases the mobility since the probability that a particle returns to the flat channel decreases with the distance to the corners. A further increase in the drift force enables transport along a secondary channel and hence increases the mobility again. This oscillatory behavior seems to continue until the external drift force completely dominates the magnetic forces.

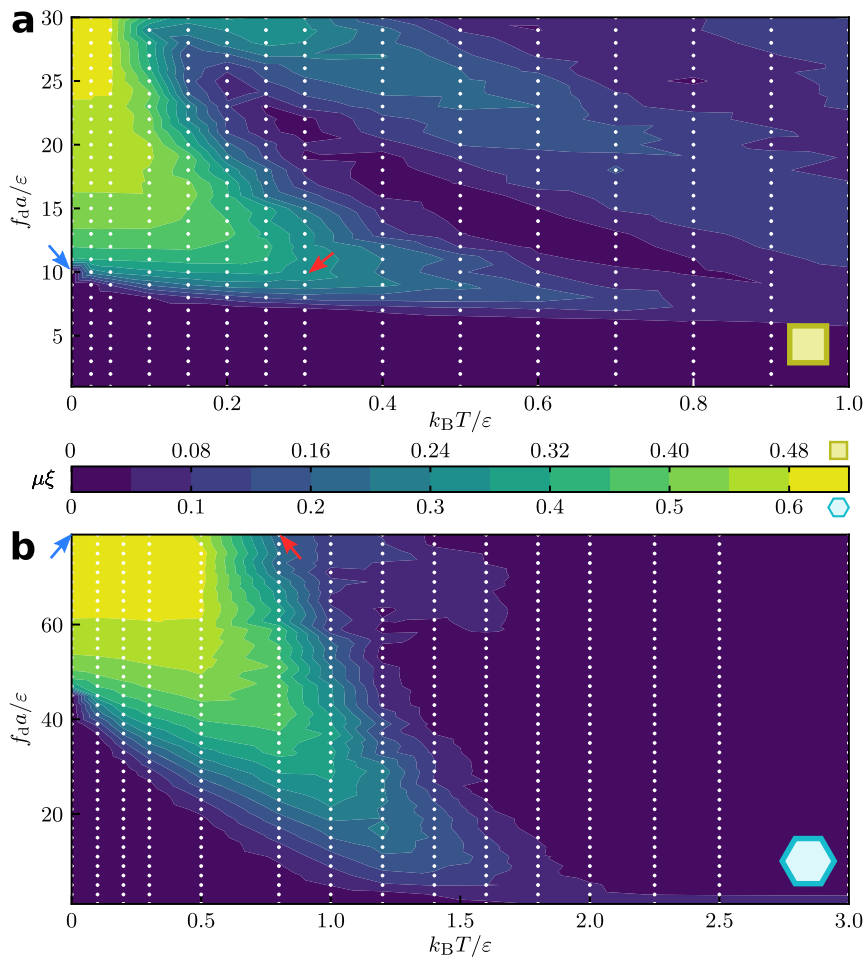

**Fig. 4 Dynamical phase diagram.** Colloidal mobility $\mu$ (see color bar) in the plane of magnitude of the drift force $f_d$ and temperature $T$ for twisted **a** square patterns at a magic angle $\alpha_m^{sq} \approx 4.24°$ and **b** hexagonal patterns at a magic angle $\alpha_m^{hex} \approx 4.41°$. The white dots indicate the selected temperatures and magnitudes of the drift force simulated to create the plots. The blue and red arrows indicate the position in the diagrams of selected state points.

In square twisted patterns, the secondary channels that are first activated for transport are those located at a distance $a/2$ from the edges for which $V_{mag}$ is maximum (the edges without flat channels), see Supplementary Movie 3. There the potential is flatter than along the secondary channels located at a distance $a$ from the flat channels and parallel to them. In contrast, in twisted hexagonal patterns, the active secondary channels are those parallel to the flat channels. Transport along the secondary channels located in edges where $V_{mag}$ is maximum does not occur simply because $\mathbf{f}_d$ is perpendicular to those channels. Supplementary Movies 4 and 5 show the motion of particles in twisted hexagonal patterns using a flat and a secondary channel, respectively. (In Supplementary Movie 4 at time $t = 11.48\tau$, a particle jumps from a secondary channel into the flat channel, where it is transported much faster.) The different type of secondary channels active for transport in square and hexagonal twisted patterns is likely the reason behind the different amplitudes of the second peak of the mobility in the dynamical phase diagrams, see Fig. 4.

**Critical force and transport for non-magic angles.** So far we have discussed the transport in patterns twisted at magic angles.

For non-magic angles, the magnetic potential is in general no longer periodic but quasiperiodic (there exist other non-magic angles for which the potential is also periodic but the unit cell contains several supercells different from each other and the transport is not optimal, see Methods and Supplementary Note 1). For non-magic angles, each supercell differs slightly from its neighbors and the flat channels are not as flat as those at a magic angle. As a result the drift force required to transport the particles increases as compared to the magic case.

For a nonperiodic magnetic potential, the critical force depends on both the initial location of the particle and the required traveled distance that is imposed a priori to calculate $f_c$. In Fig. 5 we plot $f_c$ as a function of the twist angle (scaled with the magic angle) for square and hexagonal twisted patterns. The force is calculated by averaging over a total of 100 trajectories of particles that are at time zero-initialized on different flat channels. Three data sets corresponding to the average critical force required to transport the particles a distance equivalent to the length of 10, 100, and 1000 supercells are shown. The critical force has a sharp minimum at the magic angle which gets narrower by increasing the traveled distance used to compute $f_c$. The non-smooth behavior of the critical force at non-magic angles is to be expected due to the Diophantine equations that determine the periodicity of twisted patterns, see Methods and Supplementary Note 1. We show in Supplementary Note 1 that for any angle for which the potential is nonperiodic the particles encounter at some point the most unfavorable magnetic potential along the flat channel.

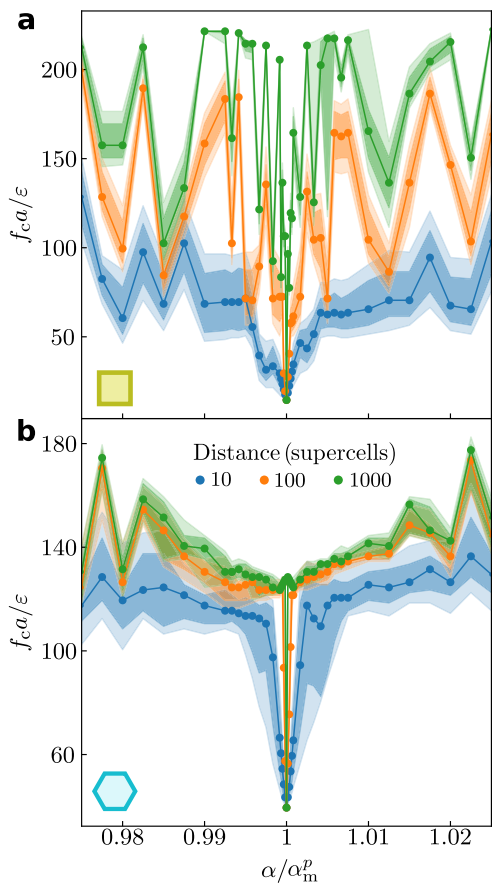

**Fig. 5 Critical force at non-magic angles.** Zero-temperature critical force $f_c$ required to transport a particle as a function of the twist angle $\alpha$ scaled with the magic angle $\alpha_m^p$ in **a** square $\alpha_m^{sq} = 6.026°$ and **b** hexagonal $\alpha_m^{hex} = 6.009°$ patterns. The different data sets show the critical force required to transport the particle at a distance equivalent to 10 (blue), 100 (orange), and 1000 (green) lengths of a supercell. Data were obtained by averaging over the trajectories of 100 particles initialized on different flat channels. The distribution of the individual measurements is illustrated by the shadow regions, which show the minimum and the maximum values of $f_c$ required to transport 50% (dark regions) and 80% (light regions) of the particles.

Hence, even the smallest deviation from a magic angle causes a substantial increase of the critical force required to transport the particles indefinitely (continuous flow).

In hexagonal twisted patterns, the continuous flow at non-magic angles and strong drift forces occur as in the magic case via the flat channels (at least for the deviations from the magic case shown in Fig. 5). At drift forces even higher than those in Fig. 5b the particles will eventually leave the flat channels in the hexagonal twisted patterns and follow the direction of $\mathbf{f}_d$. In square patterns, the particles leave the flat channels at non-magic angles and large drift forces. Hence, the continuous flow happens mostly through the entire pattern (see Supplementary Movie 6). The flat channels are deeper and narrower in hexagonal than in square twisted patterns which causes the differences in the continuous flow at non-magic angles.

We plot in Fig. 6 the average distance $d_t$ traveled by a particle located initially at the origin (fixed point of rotation) as a function of the scaled twist angle $\alpha/\alpha_m^p$. Different magic angles $\alpha_m^p$ are analysed at a fixed magnitude of the drift force. We show the zero-temperature limit (Fig. 6a, b) as well as finite temperature cases (Fig. 6c, d) for both squares (Fig. 6a, c) and hexagonal (Fig. 6b, d) twisted patterns. The value of $d_t$ is sensitive to the initial position of the particle. Nevertheless, these curves are useful to understand the physical mechanisms behind particle transport since their global characteristics are robust. The magnitude of the drift forces is above that of the critical force of the corresponding magic angle but below the magnitude required to achieve continuous flow at non-magic angles (with the exception of the red-solid lines in Fig. 6 in which $f_d$ is below the critical force at the magic angle). The required drift forces and therefore the traveled distances are higher in hexagonal than in square twisted patterns. (Recall that the critical force is larger in hexagonal than in square twisted patterns, see Fig. 3b.)

The zero-temperature limit reflects how the magnetic potential changes with the twist angle. In the inset of Fig. 6a we show how the traveled distance changes when the drift force acts during longer periods of time $t_3 > t_2 > t_1$ (see also Supplementary Movie 7). In patterns twisted at a magic angle, the number of supercells traveled is proportional to the time (provided that the drift force is above the critical one). At non-magic angles, the particles hit at some point a blocked corner/edge and stay there forever. The smaller the deviation from the magic angle is, the further from the origin this blocking occurs. As the twist angle approaches the magic angle, the region around the fixed point of rotation becomes increasingly similar to the magic case and therefore the particles travel longer distances but they eventually hit a dead end and the motion stops. In the limit of drift forces acting for an infinite period of time and comparable in amplitude to the critical force of the closest magic angle, the colloidal mobility at $T = 0$ is only different than zero if the patterns are twisted at a magic angle.

At zero-temperature, $d_t$ grows discontinuously by approaching the magic angle, Fig. 6a, b. The jumps in $d_t$ between two consecutive plateaus correspond to roughly the distance traveled across one supercell. Note that the larger scale of the plot in the hexagonal case, Fig. 6b, hinders the visualization of the plateaus but they still occur as shown in the inset.

The distance that the particles travel at the magic angle increases by decreasing $\alpha_m^p$ in square twisted patterns and it decreases in hexagonal patterns. This different behavior is, as in the case of the critical drift force, due to how the magnetic potential changes at the corners of the supercells by varying the magic angle in each type of twisted pattern.

The effect of the critical drift force for magic angles is illustrated in the data set for square patterns at $\alpha_m^{sq} \approx 4.24°$ (red line in Fig. 6a). The magnitude of the drift force, $f_d a/\varepsilon = 10$, is below the zero-temperature critical drift force for this magic angle, see Fig. 3, and therefore the transport stops completely at an angle smaller than the magic angle. However, thermal fluctuations are able to reactivate the transport (red line in Fig. 6c). At finite temperatures (Fig. 6c, d) the transport does no longer stop when a particle hits a blocking region since Brownian motion allows the particle to traverse the potential barrier (the mobilities are therefore different than zero). The temperature enhances, in general, the transport at non-magic angles. In both types of patterns, the effect is more prominent for small twist angles (blue lines in Fig. 6). However, at magic angles (and provided that the drift force is large enough to transport the particles at $T = 0$) the traveled distance is smaller at finite temperature than at $T = 0$. At magic angles, the magnetic potential is optimal for transport and Brownian motion only reduces its efficiency.

### Discussion
The patterns can be experimentally realized using e.g., exchange-bias thin magnetic films irradiated through a lithographic mask[26,27] as well as garnet films[28,29]. Using micrometer-sized colloidal particles at a distance comparable to the length of a lattice vector above these patterns[21,22] results in magnetic potential energies significantly larger than the thermal energy (at room temperature). Hence, the use of similar setups[21,22] would result in experiments on twisted patterns that are effectively close to the zero-temperature limit. The observation of the finite temperature effects discussed here should be possible by either decreasing the magnetic forces (e.g., weaker patterns and larger colloid-pattern distance) or increasing the Brownian force (using e.g., magnetic nanocolloids). As in the case of single patterns[21,22], we expect good agreement between simulations and experiments in twisted magnetic patterns.

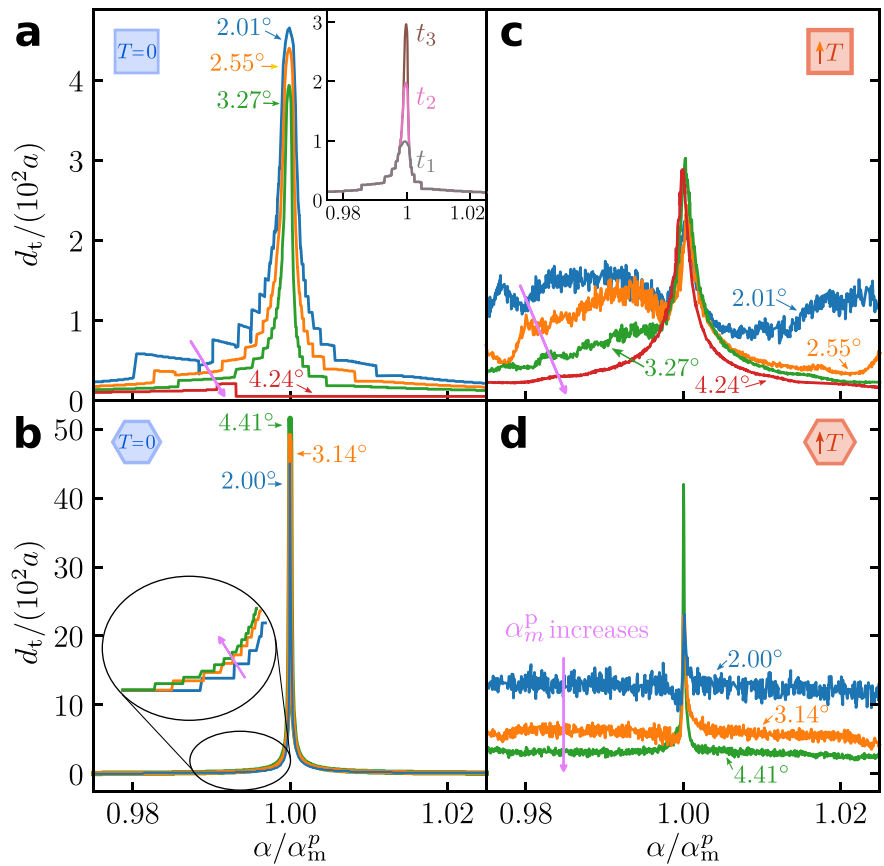

**Fig. 6 Transport at magic and non-magic angles.** Average distance traveled by a particle as a function of the twist angle $\alpha$ (scaled with the magic twist angle $\alpha_m^p$) at zero-temperature (**a**, **b**) and finite temperature (**c**, **d**) in either square (**a**, **c**) or hexagonal (**b**, **d**) twisted patterns. Data sets for several magic angles are presented, as indicated. The magic angle increases in the direction of the pink arrows. The traveled distance is obtained by averaging the motion of 100 particles located initially at the origin (axis of rotation of the patterns) and driven by a drift force (acting during $100\tau$) of magnitude $f_d a/\varepsilon = 10$ in the square (**a**, **b**) and $f_d a/\varepsilon = 80$ in the hexagonal (**c**, **d**) patterns. The temperature is $T = 0$ in (**a**) and (**b**), $k_B T/\varepsilon = 0.3$ in (**c**) and $k_B T/\varepsilon = 0.8$ in (**d**). The drift forces and temperatures used here are indicated with colored arrows in Fig. 4. The inset in (**a**) shows data for a drift force acting during $t_1 = 25\tau$, $t_2 = 50\tau$, and $t_3 = 75\tau$ (at $T = 0$ and $\alpha_m^{sq} = 3.27°$).

Regarding the drift force, colloidal particles respond to several types of external fields[30]. Gravitational[31] and electric[32] fields, as well as pressure[33] and temperature[34] gradients, are possible means to experimentally achieve such drift forces. We estimate that the strength of the Earth's gravitational field is above the critical force required to transport micrometer-sized particles in solvents with a significant particle-solvent density difference. Instead of a drift force, it would be interesting to use self-propelled active particles[35]. The twisted patterns could then be used to study and possibly tune transport properties of active Brownian particles[36], such as e.g., the polarization[37], in complex environments[38].

The colloidal transport using flat channels in twisted patterns is faster than the topologically protected transport in single patterns[21,22]. There, the particles are adiabatically driven via modulation of the orientation of the external magnetic field. In twisted patterns, topologically protected transport is also available and it can be used to e.g., initially move the colloidal particles from the inside of the supercells towards the flat channels. The availability of two types of transport mechanisms, namely flat channels and topological transport, together with the control over the flat channels offered by the twist angle are two advantages over static channels added directly to the potential in e.g., lab-on-a-chip devices.

The critical force required to transport the particles is always significantly smaller in the square than in hexagonal twisted patterns and therefore the enhanced colloidal transport is more pronounced in twisted square patterns. The effect is specially relevant approaching the limit of very small twist angles since the critical force vanishes in the case of square twisted patterns. Hence, given (i) the similarities between our colloidal system and electronic systems and (ii) the correspondence in electronic systems between flat bands in reciprocal space and flat channels in real space, it is plausible to think that twisted bilayers of two-dimensional materials with square unit cells[39–44] are promising candidates for new electronic applications.

## Methods

**Magnetic potential**. The total magnetic field at position $\mathbf{r}$ is $\mathbf{H}(\mathbf{r}) = \mathbf{H}_{p,1}(\mathbf{r}) + \mathbf{H}_{p,2}(\mathbf{r}) + \mathbf{H}_{ext}$, with $\mathbf{H}_{p,i}$ the magnetic field of pattern $i = 1, 2$ and $\mathbf{H}_{ext}$ the uniform external magnetic field, which is normal to the patterns. Hence the magnetic potential acting on a paramagnetic particle with effective volume $v_{eff}$ is

$$V_{mag} = -v_{eff}\chi\mu_0 \mathbf{H}^2(\mathbf{r}), \quad (2)$$

with $\mu_0$ the vacuum permeability and $\chi$ the particle susceptibility. We scale the particles down to effective point particles and increase their susceptibility such that $v_{\text{eff}}\chi$ remains constant. If the vertical distance between the particle and the patterns is sufficiently large (comparable or larger than the size of the unit cell of the pattern) only the first Fourier-mode of each pattern contributes significantly to the magnetic field[21,45]. Additionally, the potential is dominated by the cross-term $V_{\text{mag}} \approx -2v_{\text{eff}}\chi\mu_0 \mathbf{H}_{\text{ext}} \cdot (\mathbf{H}_{p,1} + \mathbf{H}_{p,2})$ since $\mathbf{H}_{\text{ext}}$ is uniform and much stronger than $\mathbf{H}_{p,i}$. In this limit, the system-specific parameters like the amplitudes of all magnetic fields and the distance between the patterns can be absorbed into a single constant $\varepsilon$. The potential is therefore given by

$$V_{\text{mag}} = -\chi\varepsilon a q^p \sum_{i=1}^{N} \left[ \cos\left(\mathbf{q}_i \cdot \left(\mathbf{r} - \frac{\mathbf{a}_1}{2}\right)\right) + \cos(\mathbf{R}_\alpha \mathbf{q}_i \cdot \mathbf{r}) \right], \quad (3)$$

where $N = 4, 6$ for the square and hexagonal patterns, respectively, $\mathbf{R}_\alpha$ denotes a rotation matrix by the twist angle $\alpha$ around the direction normal to the pattern, and the vectors $\mathbf{q}_i$ are given by

$$\mathbf{q}_i = q^p \begin{pmatrix} \sin(2\pi i/N) \\ \cos(2\pi i/N) \end{pmatrix}, \quad (4)$$

where the superscript p = {sq, hex} and $q^{\text{sq}} = 2\pi/a$ in the square pattern and $q^{\text{hex}} = 2\pi/(a\sin(\pi/3))$ in the hexagonal pattern. The first term in the right-hand side of Eq. (3) corresponds to the pattern shifted by $\mathbf{a}_1/2$ and the second term to the pattern twisted by an angle $\alpha$. Shifting the patterns by a different quantity also creates, in general, structures similar to flat channels. However, a shift by half a lattice vector minimizes the roughness of the magnetic potential at the flat channels and specially at the corners of the supercells by maximizing the destructive interference between the filed of both patterns (see Supplementary Note 1). Further mathematical details about the single hexagonal and square patterns can be found in previous works[21,45].

**Computer simulations**. The particle trajectories are calculated with overdamped Brownian dynamics simulations. We discretize the equation of motion, Eq. (1), using a time step $dt/\tau = 10^{-5}$ and integrate it in time via the standard Euler algorithm.

**Magic angles**. The magnetic potential of patterns twisted by an angle $\alpha$ develop moiré interference patterns at length scales roughly given by $a/(2\sin(\alpha/2))$. At magic twist angles the combined magnetic potential of both patterns becomes periodic at the length scale of the supercells. In the twisted square pattern, the potential has a checkerboard layout of two alternating supercells, which can be transformed into each other by a rotation of $\pi$ around their centers. The unit cell of a twisted square pattern is therefore twice the size of the supercell, see Fig. 1a. For hexagonal patterns twisted at a magic angle, the unit cell and the supercells coincide, see Fig. 1b. In both hexagonal and square twisted patterns there exist other twist angles (non-magic) for which the patterns are also periodic but the periodicity is recovered only after multiple supercells (which can not be transformed into each other with similarity transformations). In those patterns, the flat channels are not connected over macroscopic distances and hence the colloidal transport is not enhanced as in the magic case, see Supplementary Note 1. The mathematical condition for a magic twist angle, in which every supercell is equivalent, can be expressed as a Diophantine problem[46,47] with solution

$$\alpha_m^{\text{sq}} = 2\arcsin\left(\frac{1}{2\sqrt{k^2 + k + 1/2}}\right), \quad (5)$$

$$\alpha_m^{\text{hex}} = \arccos\left(\frac{3(2k+1)^2 - 1}{3(2k+1)^2 + 1}\right), \quad (6)$$

where $k$ is a natural number. To obtain these expressions in the hexagonal case, we adjusted the procedure by Shallcross et al.[46] to include the constraint of having identical supercells.

**Drift force**. In both hexagonal and square twisted patterns the flat channels develop along consecutive edges of the supercells. The drift force points along the bisector of the directions of the flat channels, as shown in Fig. 2. Hence,

$$\mathbf{f}_d = f_d \begin{pmatrix} \cos\alpha_d^p \\ \sin\alpha_d^p \end{pmatrix}, \quad (7)$$

where the angle $\alpha_d^p$ is given by

$$\alpha_d^{\text{sq}}(k) = \alpha_m^{\text{sq}}(k)/2 + \text{sgn}(\alpha_m^{\text{sq}}(k)) \frac{(-1)^{k+1}\pi}{4}, \quad (8)$$

$$\alpha_d^{\text{hex}}(k) = \alpha_m^{\text{hex}}(k)/2 + \text{sgn}(\alpha_m^{\text{hex}}(k)) \frac{(-1)^{k+1}\pi}{6}, \quad (9)$$

in square and hexagonal twisted patterns, and $k \in \mathbb{N}$. The factor $(-1)^{k+1}$ and the sign of the magic angle in the above expressions reflect the fact that the edges that support transport alternate from one magic angle to the next one as well as by changing the sign of the twist angle. A drift force pointing along a different direction will also induce transport provided that the force is not orthogonal to any of the directions of the flat channels.

**Critical force**. To estimate the value of the critical force we assume the following form for the mobility curves (Fig. 3a) near the transition from no-transport to transport

$$\mu(f_d) = \mu_0 \left(\frac{f_d - f_c}{f_c}\right)^\gamma, \quad (10)$$

where $\mu_0$, the critical force $f_c$ and $\gamma$ are used as fitting parameters.

**Mobility**. We define the mobility $\mu$ as the average distance traveled by the particles divided by the amplitude of the drift force

$$\mu = \frac{\langle |\mathbf{r}(t_f) - \mathbf{r}(t_i)| \rangle}{(t_f - t_i)f_d}. \quad (11)$$

We calculate the average distance $\langle |\mathbf{r}(t_f) - \mathbf{r}(t_i)| \rangle$ by initializing 100 non-interacting particles at the origin (axis of rotation) and let them travel under the influence of the drift force for a total time of $100\tau$. To eliminate the dependence on the initial conditions we average the distances traveled by the particles during the second half of the simulation, i.e., $t_i = 50\tau$ and $t_f = 100\tau$ in Eq. (11). However, to characterize the system at non-magic angles, we consider the full trajectories in Fig. 6 (such that all particles share the same initial position). Therefore, the mobilities calculated with the distances reported in Fig. 6 differ slightly from those shown in Fig. 4.

### Data availability
All the data supporting the findings are available from the corresponding author upon reasonable request.

### Acknowledgements
This work is funded by the Deutsche Forschungsgemeinschaft (DFG, German Research Foundation) under project number 440764520. Open Access funding partially provided by project DEAL. We thank the referees for their helpful comments.

### Author contributions
N.C.X.S. carried out the calculations. N.C.X.S., T.M.F., and D.d.l.H conceived and designed the idea and wrote the manuscript.

### Funding
Open Access funding enabled and organized by Projekt DEAL.

### Competing interests
The authors declare no competing interests.

### Additional information
**Supplementary information** The online version contains supplementary material available at https://doi.org/10.1038/s42005-022-00824-3.

**Correspondence** and requests for materials should be addressed to Daniel de las Heras.

**Peer review information** *Communications Physics* thanks the anonymous reviewers for their contribution to the peer review of this work.

**Reprints and permission information** is available at http://www.nature.com/reprints

**Publisher's note** Springer Nature remains neutral with regard to jurisdictional claims in published maps and institutional affiliations.